\RequirePackage{fix-cm}

\documentclass[onecolumn]{svjour3}          

\smartqed  
\usepackage{graphicx}
\usepackage{amssymb} 
 \journalname{J. Braz. Soc. Mech. Sci. Eng.}
\begin{document}

\title{THE PERTURBATION OF A TURBULENT BOUNDARY LAYER BY A TWO-DIMENSIONAL HILL \thanks{Accepted Manuscript for the Journal of the Brazilian Society of Mechanical Sciences and Engineering, v. 35, p. 337-346, 2013. The final publication is available at Springer via http://dx.doi.org/10.1007/s40430-013-0024-z}}


\author{Erick de Moraes Franklin         \and
        Guilherme Augusto Ayek 
}


\institute{Erick M. Franklin \at
              Faculty of Mechanical Engineering, University of Campinas - UNICAMP \\
              Tel.: +55-19-35213375\\
              \email{franklin@fem.unicamp.br}           
           \and
           Guilherme A. Ayek \at
              Faculty of Mechanical Engineering, University of Campinas - UNICAMP \\
              \email{guilhermeayek@hotmail.com}           
             \emph{Present address: Benteler Automotive}  
}

\date{Received: date / Accepted: date}

\maketitle

\begin{abstract}
Turbulent boundary layers over flat walls in the presence of a hill are frequently found in nature and industry. Some examples are the air flows over hills and desert dunes, but also water flows over aquatic dunes inside closed conduits. The perturbation of a two-dimensional boundary layer by a hill introduces new scales in the problem, changing the way in which velocities and stresses are distributed along the flow. When in the presence of sediment transport, the stress distribution along the hill is strongly related to bed instabilities. This paper presents an experimental study on the perturbation of a fully-developed turbulent boundary layer by a two-dimensional hill. Water flows were imposed over a hill fixed on the bottom wall of a closed conduit and the flow field was measured by Particle Image Velocimetry. From the flow measurements, mean and fluctuation fields were computed. The general behaviors of velocities and stresses are compared to published asymptotic analyses and the surface shear stress is analyzed in terms of instabilities of a granular bed.
\end{abstract}

\section{Introduction}
\label{section:introduction}

Turbulent boundary layers over flat walls are frequently found in environmental and industrial applications and, for this reason, they have been extensively studied for over a century. However, in some cases a small hill is present on the ground, perturbing the boundary layer. In nature, some examples are the air flows over hills and desert dunes, and water flows over river dunes. In industry, examples are related to flows over sand ripples and dunes in closed conduits such as petroleum pipelines, dredging lines and sewer systems.

The perturbation of the boundary layer introduces new scales in the problem, changing the way in which velocities and stresses are distributed along the flow. The new velocity and stress distributions are of importance for many engineering applications. For instance, when in the presence of sediment transport the stress distribution along the hill is an essential parameter to understand the bed instabilities \cite{Franklin_4,Franklin_5,Franklin_6}.

Over the last decades, many studies were devoted to the perturbation of a turbulent boundary layer by a low hill. Some of them, based on asymptotic methods, have improved our knowledge on the subject. In these methods, the turbulent boundary layer over a hill of small aspect ratio (height to length ratio of $O(0.1)$) is divided in a two-region structure that can be employed to determine the perturbed flow \cite{Belcher_Hunt}.  Furthermore, each of these regions is sometimes subdivided in two layers in order to correctly match each other and the boundary conditions.

Jackson and Hunt \cite{Jackson_Hunt} presented an asymptotic analysis of a turbulent boundary layer perturbed by a low hill. In their analysis, the unperturbed boundary layer was given by the law of the wall

\begin{equation}
u^+\,=\,\frac{1}{\kappa} ln \left( \frac{y}{y_0} \right) \,=\,\frac{1}{\kappa} ln(y^+) +B
\label{mean_velocity}
\end{equation}

\noindent where $\kappa\,=\,0.41$ is the von K\'arm\'an constant, $y$ is the vertical distance from the wall, $y_0$ is the roughness length, $u^+\,=\,u/u_*$ is the longitudinal component of the mean velocity normalized by the shear velocity, $u(y)$ is the longitudinal component of the mean velocity $\vec{V}$, $u_*=\rho^{-1/2}\tau_0^{-1/2}$ is the shear velocity, $\tau_0$ is the shear stress of the unperturbed flow, $\rho$ is the specific mass of the fluid, $y^+\,=\,yu_*/\nu$ is the vertical distance normalized by the viscous length, $\nu$ is the kinematic viscosity and $B$ is a constant. The second and the third terms are equivalent, the second one being generally employed in hydraulic rough regimes while the third one is employed in hydraulic smooth regimes.

Jackson and Hunt \cite{Jackson_Hunt} divided the perturbed boundary layer in two regions. The inner region, close to the bed, is a region where the turbulent vortices can adapt to equilibrium conditions with the mean flow. In this region, the time scale for the dissipation of the energy-containing eddies is much smaller than the time scale for their advection, so that this region is in local equilibrium. The local-equilibrium condition allows the use of turbulent stress models, such as the mixing-length model. In addition, as this region has a small thickness that does not change significantly along the hill, the perturbations are driven by the pressure field of the outer region.

The outer region is considered far enough from the bed so that the energy-containing vortices cannot adapt to equilibrium conditions with the mean flow: the time scale for the dissipation of the energy-containing eddies is much larger than the time scale for their advection, and the flow is not in local equilibrium. For this reason, the mean flow in this region is almost unaffected by the shear stress perturbations and a potential solution is expected at the leading order.

Jackson and Hunt \cite{Jackson_Hunt} matched these two regions and obtained a solution for the perturbation. Their composite solution shows that most of the perturbation occurs in the inner region (at each longitudinal position, the maximum of the speed-up is at approximately $1/10$ of the inner region thickness). In addition, they found that the longitudinal evolution of the perturbation has a peak upstream of the bedform crest.

Hunt et al. \cite{Hunt_1} improved the analysis of Jackson and Hunt \cite{Jackson_Hunt} by subdividing each region. They divided the inner region in two layers. In the \textit{inner surface layer}, closer to the bed, the flow is determined by pressure and shear effects (the inertial effects are negligible) and its lower part matches the boundary conditions on the bed surface. In the \textit{shear stress layer}, closer to the outer region, the flow is determined by pressure, shear, and inertial effects and its upper part matches the outer region. Hunt et al. \cite{Hunt_1} also divided the outer region in two layers. In the \textit{upper layer}, the external part of this region, the ratio between the Reynolds stress gradient and the inertial effects is very small and the flow is approximately potential. In this layer, the flow is dominated by pressure effects. In the \textit{middle layer}, lower part of this region, the shear dominates, so that the flow is inviscid, but rotational. This layer must match the shear stress layer. In addition, Hunt et al. \cite{Hunt_1} extended the analysis to three-dimensional hills.

The results obtained by Hunt et al. \cite{Hunt_1} are in agreement with that of Jackson and Hunt \cite{Jackson_Hunt}. Hunt et al. \cite{Hunt_1} also showed that the maximum of the perturbation velocity occurs in the shear stress layer and that near the surface the relative increase of the surface stress is greater than that of velocity.

Weng et al. \cite{Weng} further developed the works of Jackson and Hunt \cite{Jackson_Hunt} and Hunt et al. \cite{Hunt_1}. They computed the velocity perturbations until the second order, obtaining a smoother matching, and applied the results to forms with higher aspect ratios. Their proposed expressions for the surface stresses, at the first order, are largely employed.

Sauermann \cite{Sauermann_2} and Kroy et al. \cite{Kroy_A,Kroy_C} simplified the results of Weng et al. \cite{Weng} for the surface stress and obtained an expression containing only the dominant physical effects of the perturbation, making clear the reasons for its upstream shift. For a hill with local height $h$ and a length $2L$ between the half-heights (total length $\approx 4L$), they showed that the perturbation of the longitudinal shear stress (dimensionless) is

\begin{equation}
\hat{\tau}_{x}=A\left(\frac{1}{\pi}\int{\frac{\partial_{x}h}{x-\xi}d\xi}\,+\,B_e\partial_{x}h\right)
\label{stress_pert_reel}
\end{equation}

\noindent where $\xi$ is an integration variable and $A$ and $B_e$ are considered as constants, as they vary with the logarithm of $L/y_0$. Variations in three orders of magnitude of $L/y_0$,  $L/y_0=10^3$,  $L/y_0=10^4$ and $L/y_0=10^5$, give $A=4.0$, $A=3.6$ and $A=3.3$ and $B_e=0.63$, $B_e=0.46$ and $B_e=0.36$, respectively. The first term in the parentheses, the convolution product, is symmetric, similar to the potential solution of the flow perturbation by a hill. It comes from the pressure perturbations caused by the hill on the outer region. The second term in the parentheses, which takes into account the local slope, is anti-symmetric. It comes from the nonlinear inertial terms of the turbulent flow and can be seen as a second order correction of the potential solution, with minor changes in the magnitude of the first order solution, but causing an upstream shift. In the Fourier space, Eq. \ref{stress_pert_reel} may be written as (dimensionless)

\begin{equation}
\hat{\tau}_{k}=Ah(|k|+iB_ek)
\label{stress_pert_fourier}
\end{equation}

\noindent where $k=2\pi\lambda^{-1}$ is the longitudinal wavenumber ($\lambda$ is the wavelength) and $i$ is the imaginary number. If the perturbation is supposed small compared to a basic flow, the fluid flow over the bed can be written as a basic flow, unperturbed, plus a flow perturbation. For the shear stress on the bed surface

\begin{equation}
\tau\,=\,\tau_{0}(1\,+\,\hat{\tau})
\label{stress_total}
\end{equation}

\noindent In the case of loose granular beds, the longitudinal evolution of the shear stress determines if the fluid flow is an unstable mechanism, so that Eqs. \ref{stress_pert_reel} to \ref{stress_total} are of importance for stability analyses of sand ripples and dunes \cite{Franklin_4,Franklin_5,Franklin_6,Franklin_8}.

\begin{sloppypar}

Formally, asymptotic methods are applied to hills with aspect ratios of $h_{max}\left( 4L\right)^{-1}<0.05$ \cite{Jackson_Hunt,Hunt_1,Sykes}, where the total length of the bedform is approximately $4L$ and $h_{max}$ corresponds to its maximum height. Carruthers and Hunt \cite{Carruthers_Hunt} showed that reasonable results are obtained when applied to slopes up to $h_{max}\left( 4L\right)^{-1}=0.3$ (note that the aspect ratio of dunes is $h_{max}\left( 4L\right)^{-1}=O(0.1)$). In particular, when $h_{max}\left( 4L\right)^{-1}=O(0.1)$, the obtained equations shall be applied to an envelope formed by the bedform and the recirculation bubble \cite{Weng}.

Recently, Franklin and Charru \cite{Franklin_9} and Charru and Franklin \cite{Franklin_10} studied the isolated three-dimensional dunes, known as barchans, in the specific case of closed-conduit water flows. In particular, the evolution of the shear stress along the symmetry plane of the dune was investigated. Different from the aeolian case, the authors found that the surface shear stress is not shifted upstream of the dune crest. If this is true, the liquid flow is not the unstable mechanism and the formation of aquatic barchans cannot be understood. The absence of an upstream shift was not explained by the authors, the reason being probably linked to the flow three-dimensionality.

\end{sloppypar}

This paper presents an experimental study of the perturbation of a turbulent boundary layer by a two-dimensional hill. A closed-conduit water flow was imposed over a triangular ripple and the flow was measured by PIV (Particle Image Velocimetry). From the flow measurements, the mean velocities and the fluctuations were computed, so that the shear stress over the ripple could be determined. The general behaviors of velocities and stresses are compared to published asymptotic analyses and the surface stress over the ripple is discussed in terms of bed instabilities.

Section \ref{section:setup} presents the experimental set-up and Section \ref{section:results} presents and discusses the experimental results. The conclusion section follows.

\section{Nomenclature}
\label{section:nomenclature}

$A$ = constant\\
$B$ = constant\\
$B_e$ = constant\\
$f$ = Darcy friction factor\\
$g$	= acceleration of gravity, $ms^{-2}$\\
$H$	= channel height, $m$\\
$h$	= hill's local height, $m$\\
$H_{eff}$ = distance from the PVC bed to the top wall, $m$\\
$i$	= imaginary number\\
$k$	= wavenumber, $m^{-1}$\\
$L$	= longitudinal distance between the crest and the position where the local height is half of its maximum value, $m$\\
$Q$ = water flow rate, $m^3/h$\\
$Re$ = channel Reynolds number, $Re=\overline{U}2H_{eff}/\nu$\\
$\overline{U}$ = cross-section mean velocity of the fluid, $m/s$\\ 
$u$	= longitudinal component of the mean fluid velocity, $ms^{-1}$\\
$u'$ = longitudinal component of the velocity fluctuation, $ms^{-1}$\\
$u_*$ = shear velocity, $ms^{-1}$\\
$u^+$ = dimensionless velocity, $u^+\,=\,u/u_*$\\
$-\overline{u'v'}$ = $xy$ component of the Reynolds shear stress, $(m/s)^2$\\ 
$\vec{V}$ = mean fluid velocity, $ms^{-1}$\\
$v$	= vertical component of the mean fluid velocity, $ms^{-1}$\\
$v'$ = vertical component of the velocity fluctuation, $ms^{-1}$\\
$x$ = longitudinal coordinate, $m$\\
$y$ = vertical coordinate, $m$\\
$y_d$ = displaced vertical coordinate, $m$\\
$y_0$ = roughness length, $m$\\
$y^+$ = dimensionless vertical coordinate, $y^+\,=\,yu_*/\nu$\\

\textbf{Greek symbols}\\
$\kappa$ = von K\'arm\'an constant\\
$\lambda$ = wavelength, $m$\\
$\nu$ = kinematic viscosity, $m^2/s$\\
$\rho$ = specific mass of the fluid, $kg/m^3$\\
$\tau$ = shear stress on the bed, $N/m^2$\\ 
$\xi$ = integration variable, $m$\\

\textbf{Subscripts}\\
$k$ = relative to the Fourier space\\
$x$ = relative to the real space\\
$0$ = relative to the flat wall (except in $y_0$)\\

\textbf{Superscripts}\\
$\hat{}$ = relative to the perturbation\\
$'$ = fluctuation\\

\section{Experimental set-up}
\label{section:setup}

The experimental device consisted of a water reservoir, a progressive pump, a flow straightener, a $5m$ long transparent channel of rectangular cross section ($160mm$ wide by $50mm$ high), a settling tank and a return line, so that the water flowed in a closed loop. 

\begin{figure}
  \begin{center}
    \includegraphics[width=0.98\columnwidth]{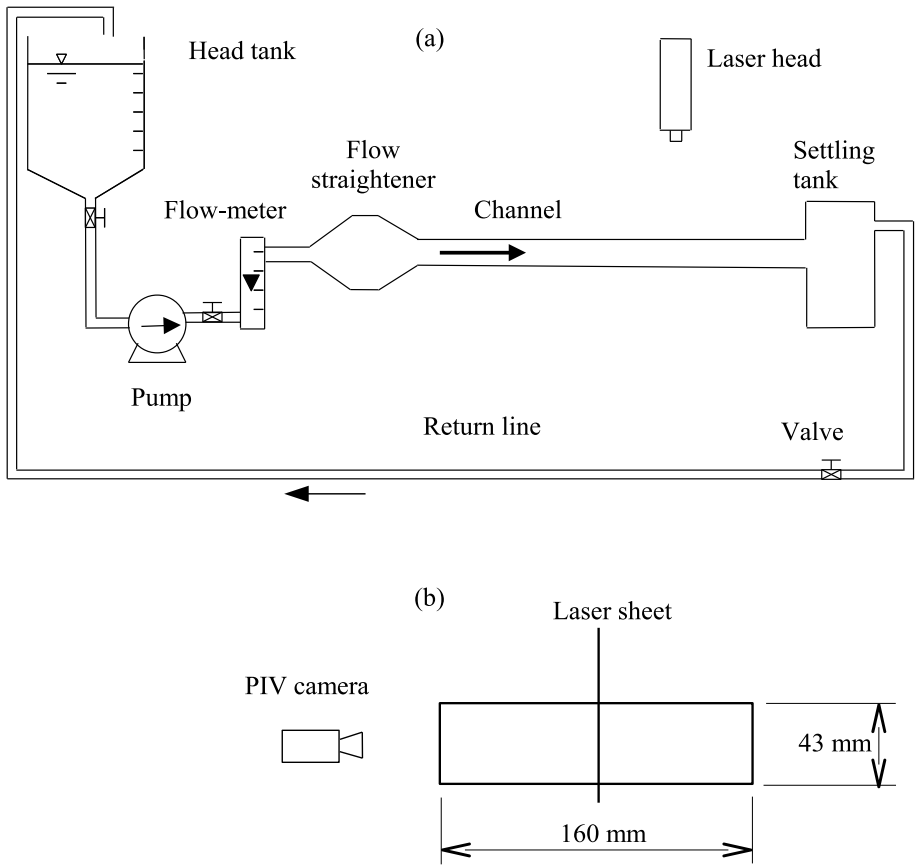}
    \caption{Layout of the experimental device: (a) side-view; (b) cross section.}
    \label{fig:schema}
  \end{center}
\end{figure}

The flow straightener was placed at the channel inlet and consisted of a divergent-convergent nozzle filled with $d=3mm$ glass spheres, whose function was to homogenize the water flow profile. The channel test section was $1m$ long and started at $40$ hydraulic diameters ($3m$) downstream of the channel inlet. There was another $1m$ long section connecting the test section exit to a settling tank and the return line. A layout of the experimental device is presented in Fig. \ref{fig:schema}

\begin{figure}
  \begin{center}
    \includegraphics[width=0.98\columnwidth]{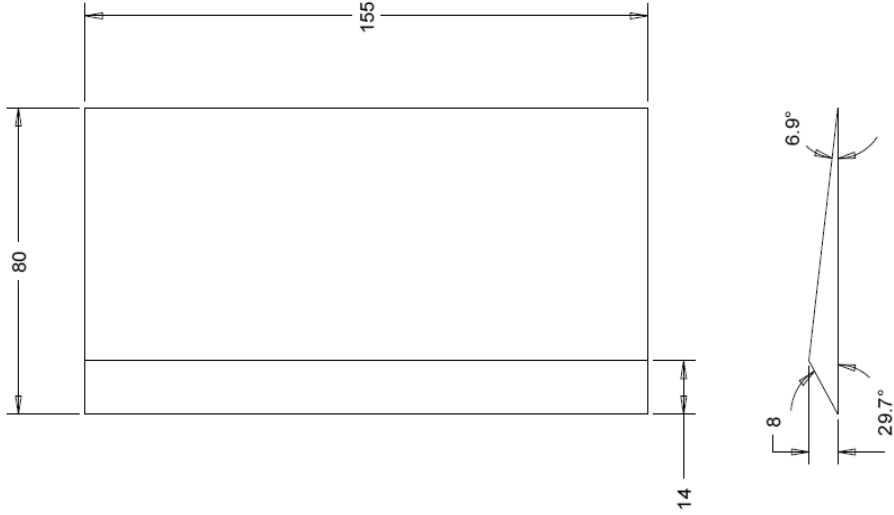}
    \caption{Bedform of triangular profile employed as a model ripple (in the figure, the flow is downwards).}
    \label{fig:ripple}
  \end{center}
\end{figure}

In order to reduce the height of the channel, PVC plates of $7mm$ thickness were inserted in the channel, covering its entire bottom. To model a two-dimensional ripple, a small bedform of triangular shape was fixed on a PVC plate in the test section. The triangular bedform had the same scales as the aquatic ripples \cite{Franklin_4} and some closed-conduit dunes \cite{Franklin_5,Franklin_2}: $80mm$ long, $8mm$ height, an upwind angle of $6.9^o$ and a lee-side angle of $29.7^o$ (close to the repose angle). The triangular bedform, of PVC, was painted in black (tarnished) in order to minimize undesirable reflections. Figure \ref{fig:ripple} presents the dimensions of the triangular bedform.

\begin{sloppypar}
The employed flow rates varied between $5$ and $10m^3/h$. They were controlled by changing the excitation frequency of the pump and measured by an electromagnetic flow-meter. These flow rates corresponded to cross-section mean velocities $\overline{U}$ within $0.20$ and $0.40m/s$ and to Reynolds numbers $Re=\overline{U}2H_{eff}/\nu$ within $1.7\cdot 10^4$ and $3.5\cdot 10^4$, where $H_{eff}$ is the distance from the surface of the PVC plates to the top wall of the channel. For different flow rates, measurements were performed without and with the ripple in the closed conduit.
\end{sloppypar}

Particle Image Velocimetry was employed to obtain the instantaneous velocity fields of the water stream. The employed light source was a dual cavity Nd:YAG Q-Switched laser, capable to emit at $2\,\times\,130mJ$ at a $15Hz$ pulse rate. The power of the laser beam was fixed at $80\%$ of the maximum power in order to assure a good balance between the image contrasts and undesirable reflection from the channel walls. Suspension of particles already present in the city (tap) water and $10\,\mu m$ hollow glass beads ($S.G.=1.05$) were employed as seeding particles.

\begin{sloppypar}	
The PIV images were captured by a $7.4\mu m \times 7.4\mu m$ ($px^2$) CCD (charge coupled device) camera with a spatial resolution of $2048px\,\times\,2048px$ and acquiring pairs of images at $4Hz$. The total field employed was of $140mm\,\times\,140mm$, corresponding to a magnification of $0.1$, and the employed interrogation area was of $8px\,\times\,8px$, corresponding to $60\mu m \times 60\mu m$ in the CCD. Considering diffraction, the diameter of the seeding particles was around $4 \mu m$ in the CCD. The computations were made with $50\%$ of overlap, corresponding to $512$ interrogation areas and to a spatial resolution of $0.27mm$. The minimum distance from the wall from which measurements were considered valid was of the order of $1mm$.

\begin{figure}
  \begin{center}
    \includegraphics[width=0.95\columnwidth]{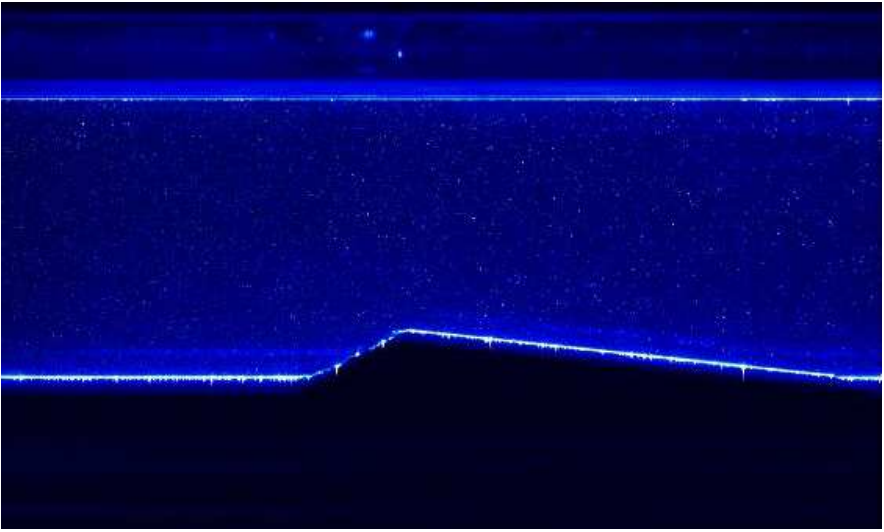}
    \caption{Image of a PIV experiment in the presence of a ripple. In this image, the flow is from right to left.}
    \label{fig:piv}
  \end{center}
\end{figure}

Each experimental run acquired $1000$ pairs of images for the tests with the ripple and $500$ pair of images for the tests without the ripple, from which the fields of instantaneous velocity, of time-averaged velocity and of the velocity fluctuations were computed in fixed Cartesian grids by the PIV controller software. MatLab scripts were written to post-process these fields (spatio-temporal averaged profiles, shear velocities, stresses on the ripple coordinate system, longitudinal evolutions, etc.).  Figure \ref{fig:piv} presents an example of PIV image for the experiments with a ripple.
\end{sloppypar}

\section{Results}
\label{section:results}

\subsection{Channel flow}
\label{subsection:channel}

The water flow was first measured in the absence of the ripple, corresponding then to a turbulent, fully-developed channel flow. This case is indicated in the following by the subscript $0$. For each test, the instantaneous fields were time-averaged and the fluctuation fields (second-order moments) were computed and time-averaged. As the flow was fully developed, the time-averaged fields were space-averaged in the longitudinal direction. With this procedure, vertical profiles of the mean velocities and of second-order moments were obtained taking advantage of the spatial resolution of the PIV equipment.

\begin{figure}
\begin{center}
	\begin{tabular}{c}
	\includegraphics[width=0.80\columnwidth]{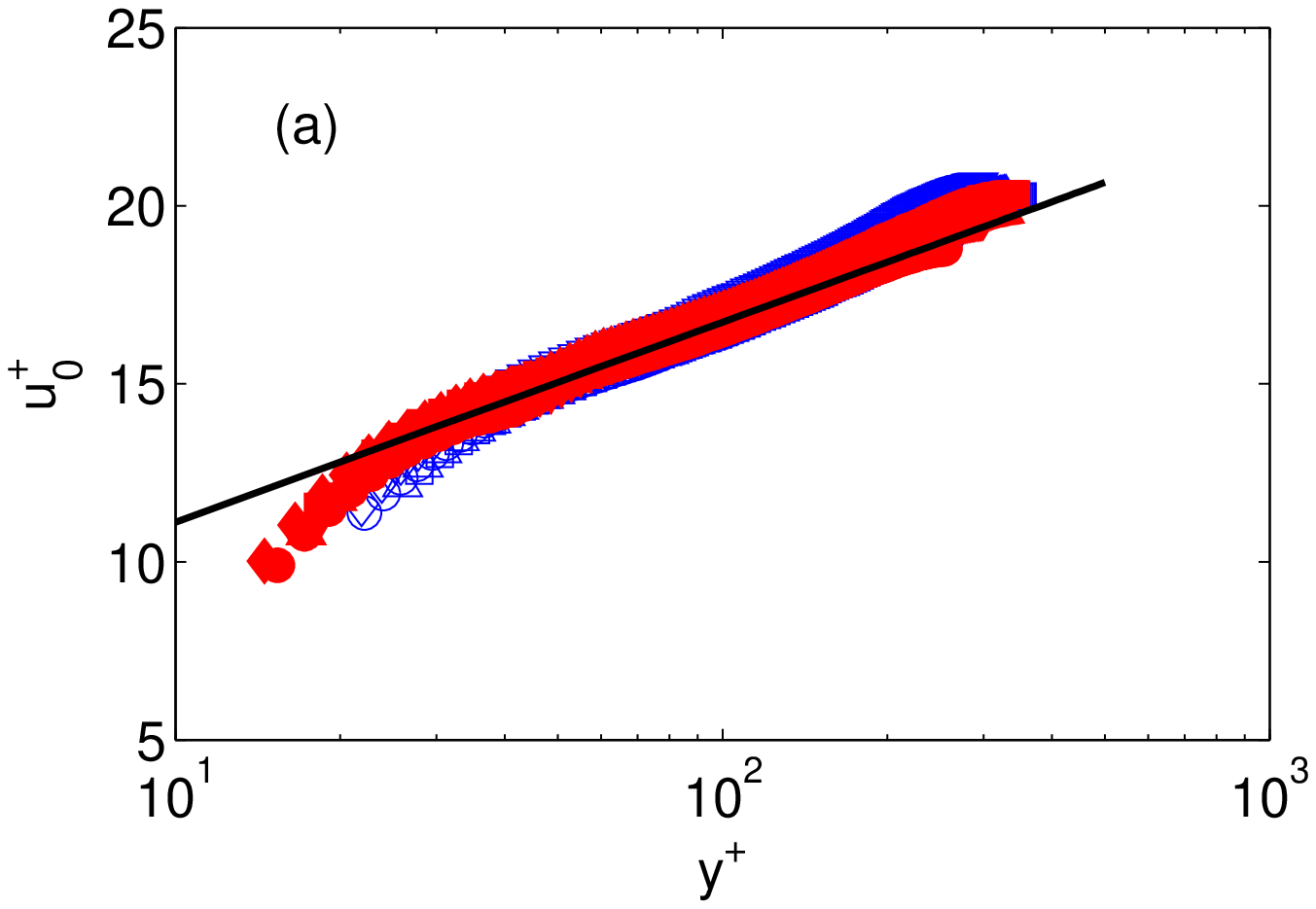}\\
	\includegraphics[width=0.80\columnwidth]{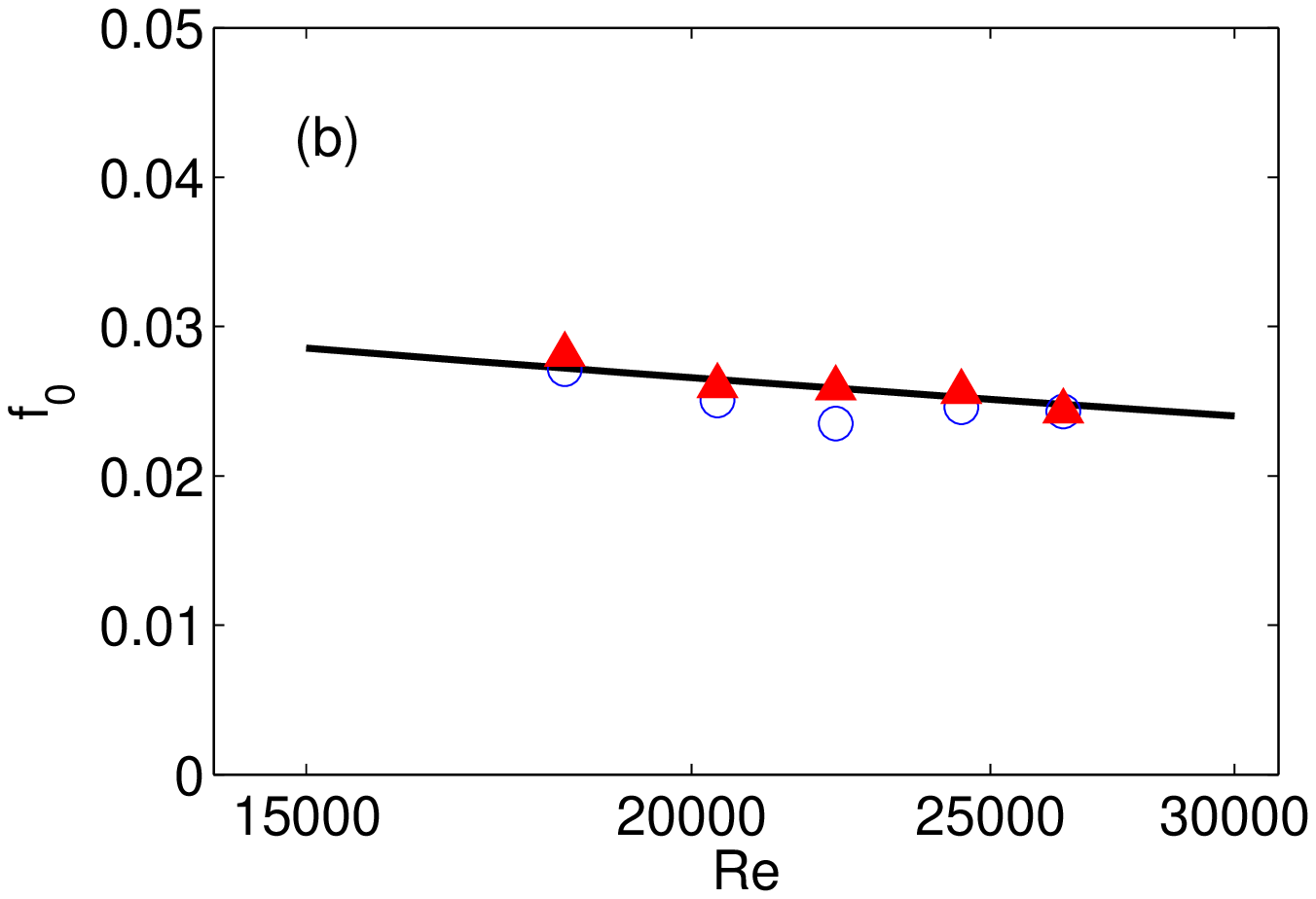}
	\end{tabular} 
\end{center}
    \caption{Mean channel flow: (a) velocity profiles. The employed symbols are listed in Table \ref{table_channel}; (b) friction factor.}
    \label{fig:mean_channel}
\end{figure}

\begin{sloppypar}	
For a fully-developed turbulent flow in a two-dimensional channel, only the longitudinal component of the mean velocity is present and Eq.\ref{mean_velocity} is valid. The shear velocity $u_{*,0}$ for each Reynolds number was then determined by fitting the experimental data in the logarithmic region ($70<y^+<200$) with Eq. \ref{mean_velocity} for a hydraulic smooth regime ($B_0=B\approx 5.5$) \cite{Schlichting_1}. The obtained values of $u_{*,0}$ and $B_0$ as well as the symbols employed in Fig. \ref{fig:mean_channel}a are presented in table \ref{table_channel}.
\end{sloppypar}	

\begin{table}[htbp]
\caption{Computed shear velocity $u_{*,0}$ and constant $B_0$ for each water flow rate $Q$.}
\label{table_channel}
\begin{center}
\begin{tabular}{|c|c|c|c|c|c|}
	\hline
	$Q$ & $\overline{U}$ & Re & Symbol & $B_0$ & $u_{*,0}$\\
	$m^3/h$ & $m/s$ & $\cdots$ & $\cdots$ & $\cdots$ & $m/s$\\
	\hline
	$5.0$ & $0.20$ & $1.8 \cdot 10^4$ & $\bigcirc$ & $5.3$ & $0.0117$\\
	\hline
	$5.6$ & $0.23$ & $2.0 \cdot 10^4$ & $\Diamond$ & $5.7$ & $0.0126$\\
	\hline
	$6.1$ & $0.25$ & $2.2 \cdot 10^4$ & $\bigtriangledown$ & $6.0$ & $0.0134$\\
	\hline
	$6.8$ & $0.27$ & $2.4 \cdot 10^4$ & $\bigtriangleup$ & $5.2$ & $0.0150$\\
	\hline
	$7.3$ & $0.29$ & $2.6 \cdot 10^4$ & $\square$ & $5.2$ & $0.0161$\\
	\hline
\end{tabular}
\end{center}
\end{table}

Figure \ref{fig:mean_channel}a presents the log-normal profiles of the mean velocities for different Reynolds numbers. The abscissa is in logarithmic scale and represents the vertical distance from the channel walls (bottom or top) normalized by the viscous length, $y^+$. The ordinate is in linear scale and corresponds to the mean velocities normalized by the shear velocity, $u^+_0$. Given the logarithmic scales of $y^+$, the profiles for each flow rate are depicted in two parts: one from the bottom wall until the channel axis of symmetry, represented by the open symbols; and the other from the top wall until the channel axis of symmetry, represented by the filled symbols.

Figure \ref{fig:mean_channel}a shows that the profiles follow the law of the wall, with a well defined logarithmic region. In addition, the profiles for different Reynolds numbers, for both the bottom and the top walls, are perfectly superposed. This justifies \textit{a posteriori} the hydraulic smooth assumption and indicates that the shear velocities were determined correctly.

Based on the shear and the cross-section velocities, the Darcy friction factor was computed as

\begin{equation}
f_0=8\left(\frac{u_{*,0}}{\overline{U}}\right)^2
\label{darcy}
\end{equation}

The friction factor $f_0$ as a function of the Reynolds number $Re$ is presented in Fig. \ref{fig:mean_channel}b. The open symbols correspond to $f_0$ on the bottom wall, the filled symbols to $f_0$ on the top wall and the continuous line corresponds to the Blasius correlation for smooth walls $f_0=0.316Re^{-0.25}$ \cite{Schlichting_1}. The computed values of $f_0$ are in good agreement with the Blasius correlation, corroborating the values obtained for the shear velocity $u_{*,0}$.

\begin{figure}
  \begin{center}
    \includegraphics[width=0.80\columnwidth]{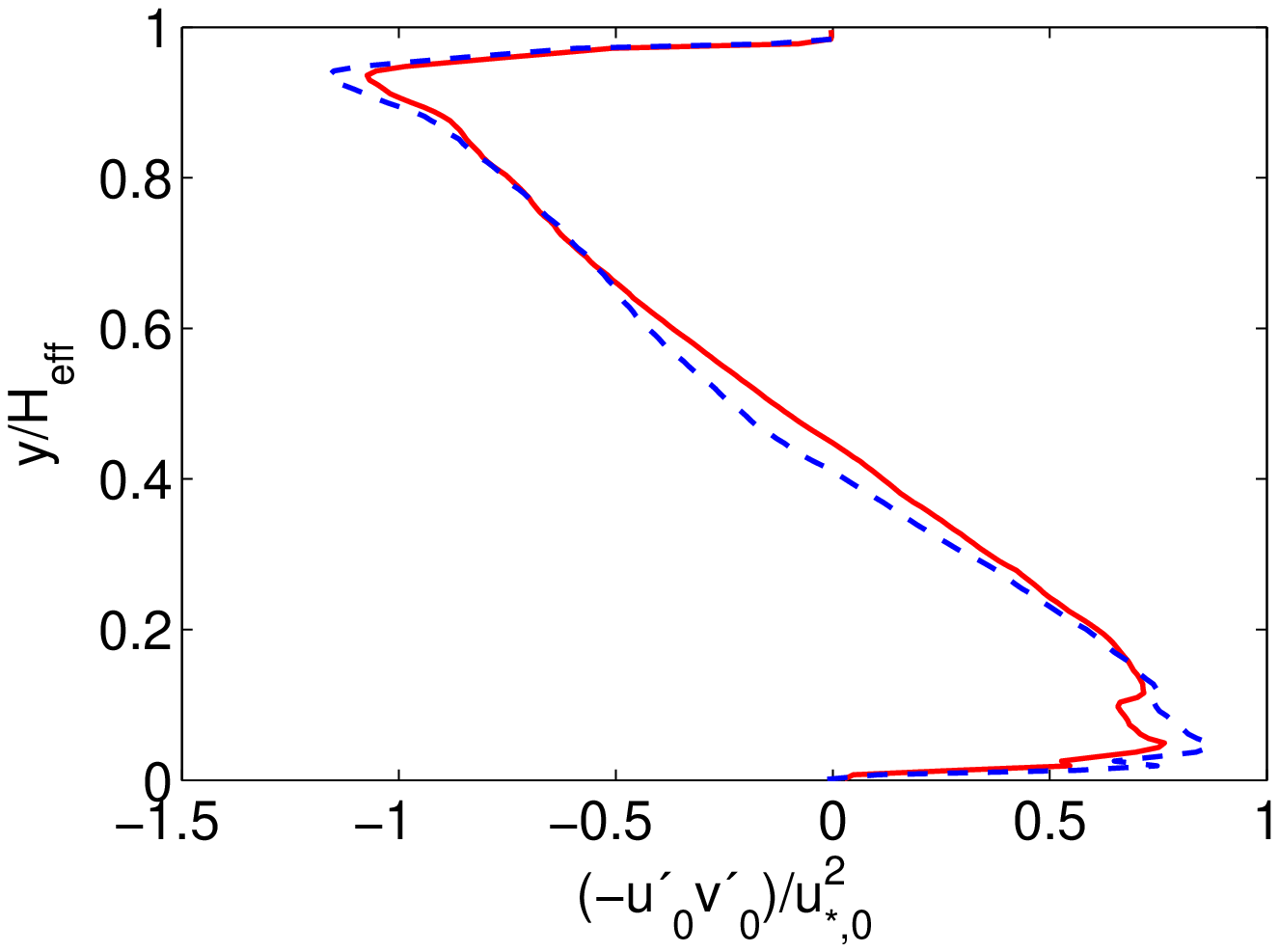}
    \caption{Profiles of the $xy$ component of the Reynolds stress in dimensionless form: $y/H_{eff}$ versus $-\overline{u'_0v'_0}/(u_{*,0}^2)$. The continuous line corresponds to $Re=1.8 \cdot 10^4$ ($\overline{U}=0.20m/s$) and the dashed line corresponds to $Re=2.6 \cdot 10^4$ ($\overline{U}=0.30m/s$)}
    \label{fig:reynolds_canal}
  \end{center}
\end{figure}

Higher uncertainties are involved in the obtainment of the Reynolds stress component $-\overline{u'_0v'_0}$. In order to decrease the noise, the $-\overline{u'_0v'_0}$ profiles were averaged by a sliding window process over the closest $9$ points. Figure \ref{fig:reynolds_canal} presents the Reynolds stress profiles in dimensionless form: $y/H_{eff}$ versus $-\overline{u'_0v'_0}/(u_{*,0}^2)$. Two different flow conditions are presented, the continuous line corresponding to $Re=1.8 \cdot 10^4$ ($\overline{U}=0.20m/s$) and the dashed line to $Re=2.6 \cdot 10^4$ ($\overline{U}=0.30m/s$). Although in the presence of more noise, the $-\overline{u'_0v'_0}$ profiles are as expected for channel flows \cite{Schlichting_1}.

In summary, the law of the wall and the Blasius correlation are valid for the turbulent flow in the test section and therefore can be used to estimate the unperturbed flow.

\subsection{Perturbed flow}
\label{subsection:ripple}

The water flow was measured over the model ripple for $Q=5m^3/h$, $7.5m^3/h$ and $10m^3/h$, which correspond to $\overline{U}=0.20m/s$, $0.30m/s$ and $0.40m/s$ and to $Re=1.8\cdot 10^4$, $2.6\cdot 10^4$ and $3.5\cdot 10^4$. In the following, the perturbed flow is analyzed and compared to the channel flow.

\begin{figure}
  \begin{center}
    \includegraphics[width=0.98\columnwidth]{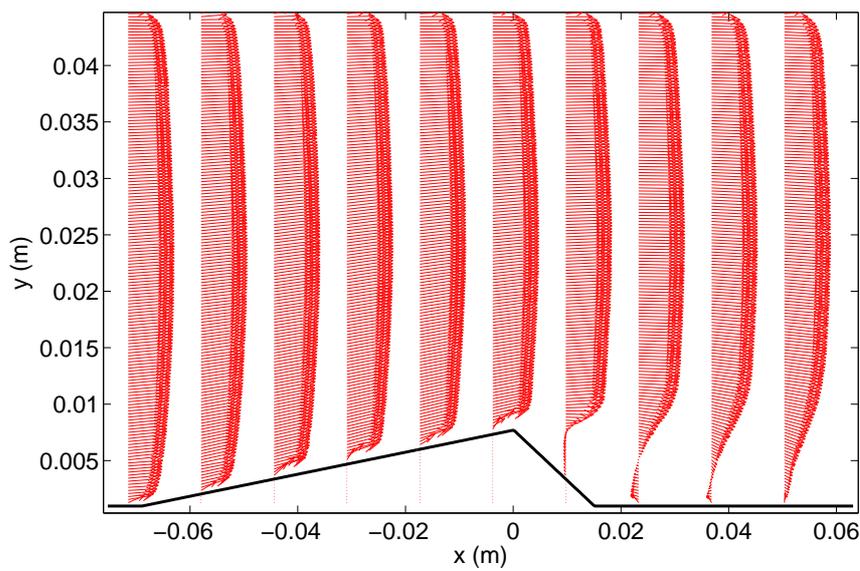}
    \caption{Some profiles of the perturbed mean velocities $\vec{V}$ over the ripple. The flow is from left to right and $Re=3.5\cdot 10^4$.}
    \label{fig:quiver_umean}
  \end{center}
\end{figure}

Figure \ref{fig:quiver_umean} presents some mean velocity profiles $\vec{V}(y)$ over the ripple. A total of $512$ profiles were obtained with the spatial resolution of the employed PIV device, however Fig. \ref{fig:quiver_umean} presents only $10$ of them. This smaller number of profiles allows a better visualization of the flow field. In this figure, as well as in the following ones, the origin of the longitudinal coordinate was set to coincide with the ripple crest. 

The main characteristics of the mean flow can be obtained from Fig. \ref{fig:quiver_umean}, which shows that the water stream is deflected by the ripple. Close to the ripple surface, the vertical component of the mean flow $v$ is no longer negligible. Upstream of the crest, $v$ is directed upwards. Downstream of the crest, the water flow detaches and a recirculation bubble is formed: $v$ is directed upwards just downstream of the crest and downwards some distance farther. Far from the ripple surface, the values of $v$ are negligible. However, although $v\approx 0$, the longitudinal component $u$ is accelerated in this region, as expected from the mass conservation.

To proceed with a boundary-layer analysis, different velocity profiles must have as reference the solid surface, i.e., the vertical position where $\vec{V}=0$. In the upper region, far from the ripple surface, the channel walls are suitable references and the vertical coordinate $y$ can be employed. In the lower region, close to the ripple, the ripple surface is the reference and therefore the displaced vertical coordinate $y_d$, given by Eq. \ref{displaced}, is the proper coordinate 

\begin{equation}
y_d\,=\,y-h
\label{displaced}
\end{equation}

\begin{figure}
\begin{center}
	\begin{tabular}{c}
	\includegraphics[scale=0.5]{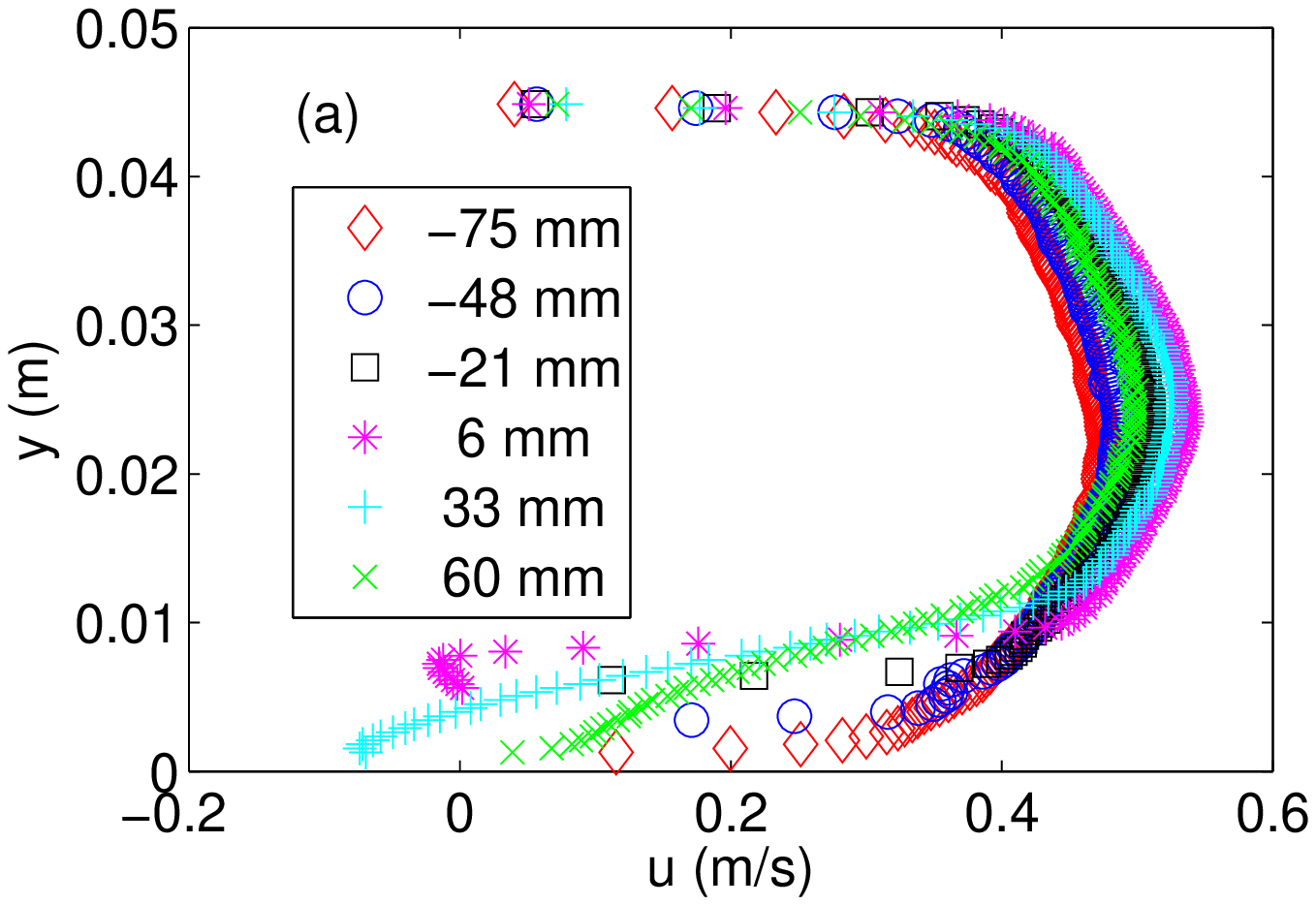}\\
	\includegraphics[scale=0.5]{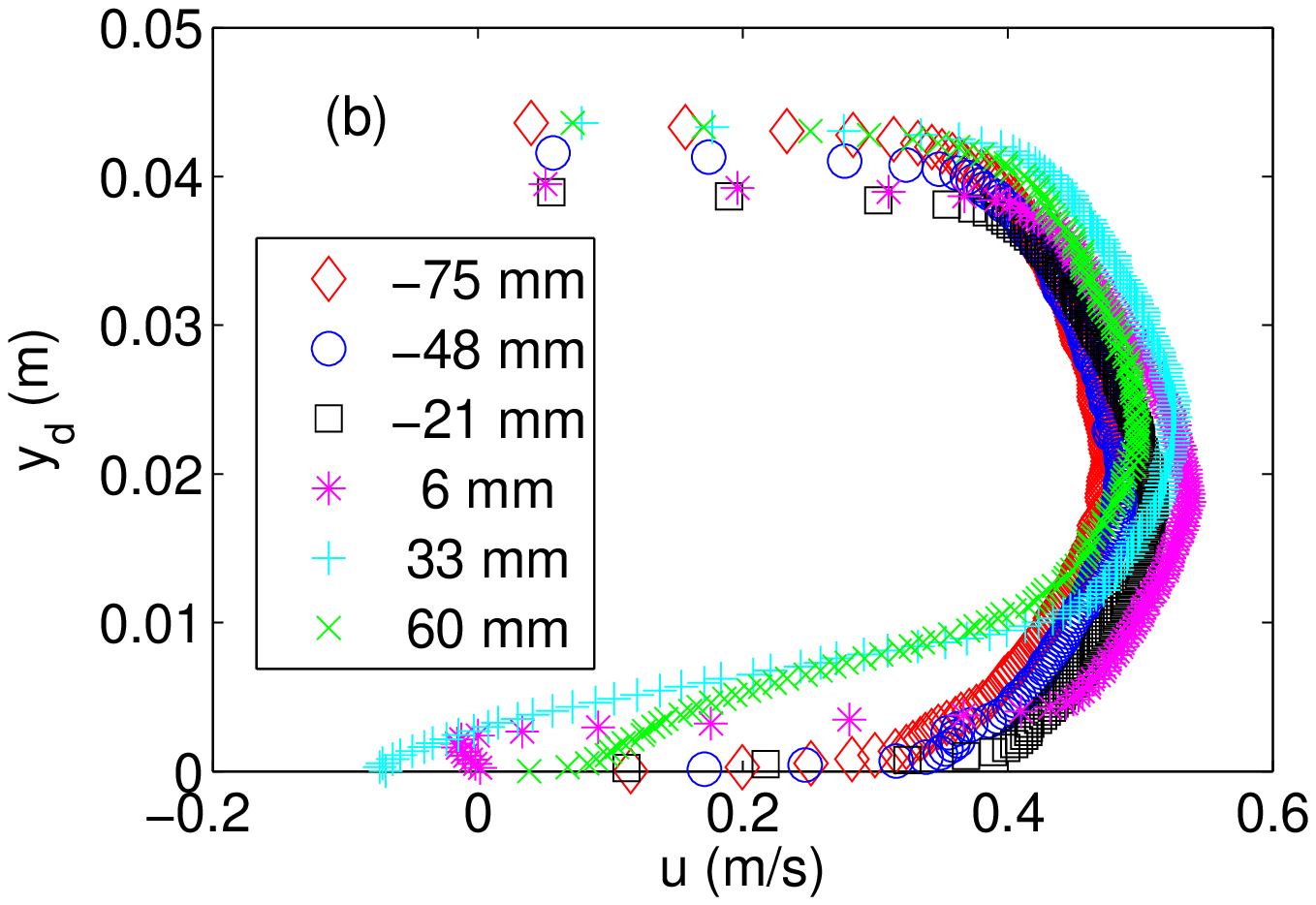}
	\end{tabular} 
\end{center}
\caption{(a) Vertical coordinate $y$ versus the longitudinal mean velocity $u$. (b) Displaced vertical coordinate $y_d$ versus the longitudinal mean velocity $u$. $Re=3.5\cdot 10^4$.}
\label{fig:umean_ripple}
\end{figure}

\begin{sloppypar}
Figure \ref{fig:umean_ripple} presents the longitudinal component of some mean velocity profiles for different longitudinal positions. Each employed symbol corresponds to the longitudinal position indicated in the legends. Figure \ref{fig:umean_ripple}(a) presents $y$ versus $u$, being suitable for the analysis of the core flow and upper wall regions, called here \textit{upper region}. In the upper region, $u$ increases as the flow approaches the position of the ripple crest ($x=0\,m$), reaching a maximum near $x=0\,m$. The longitudinal acceleration is due to the restriction caused by the combined effects of the ripple surface and the recirculation bubble so that the maximum of $u$ does not occur exactly at $x=0\,m$. Downstream of $x\approx 0\,m$, $u$ decelerates.
\end{sloppypar}

Figure \ref{fig:umean_ripple}(b) presents $y_d$ versus $u$, being suitable for the analysis of the region close to the ripple surface, called here \textit{lower region}. We observe in this region an increase of $u$ as the flow approaches the ripple crest, however the acceleration is stronger then in the upper region. A great part of the perturbation is confined in the lower region (this will become clearer later), therefore the maximum of the longitudinal velocity does not coincide with that of the upper region. Downstream of the crest, the flow detaches and $u$ has negative values.

The longitudinal evolution of the perturbed flow is of importance to understand the formation of sand ripples. For the growth of ripples, the fluid flow must be an unstable mechanism, which means that the surface shear stress caused by the fluid must reach its maximum upstream of the bedform crests \cite{Franklin_4,Franklin_5,Franklin_6,Franklin_8}. Therefore, in the following we focus our attention on the region upstream of the ripple crest.

\begin{figure}
\begin{center}
	\begin{tabular}{c}
	\includegraphics[scale=0.5]{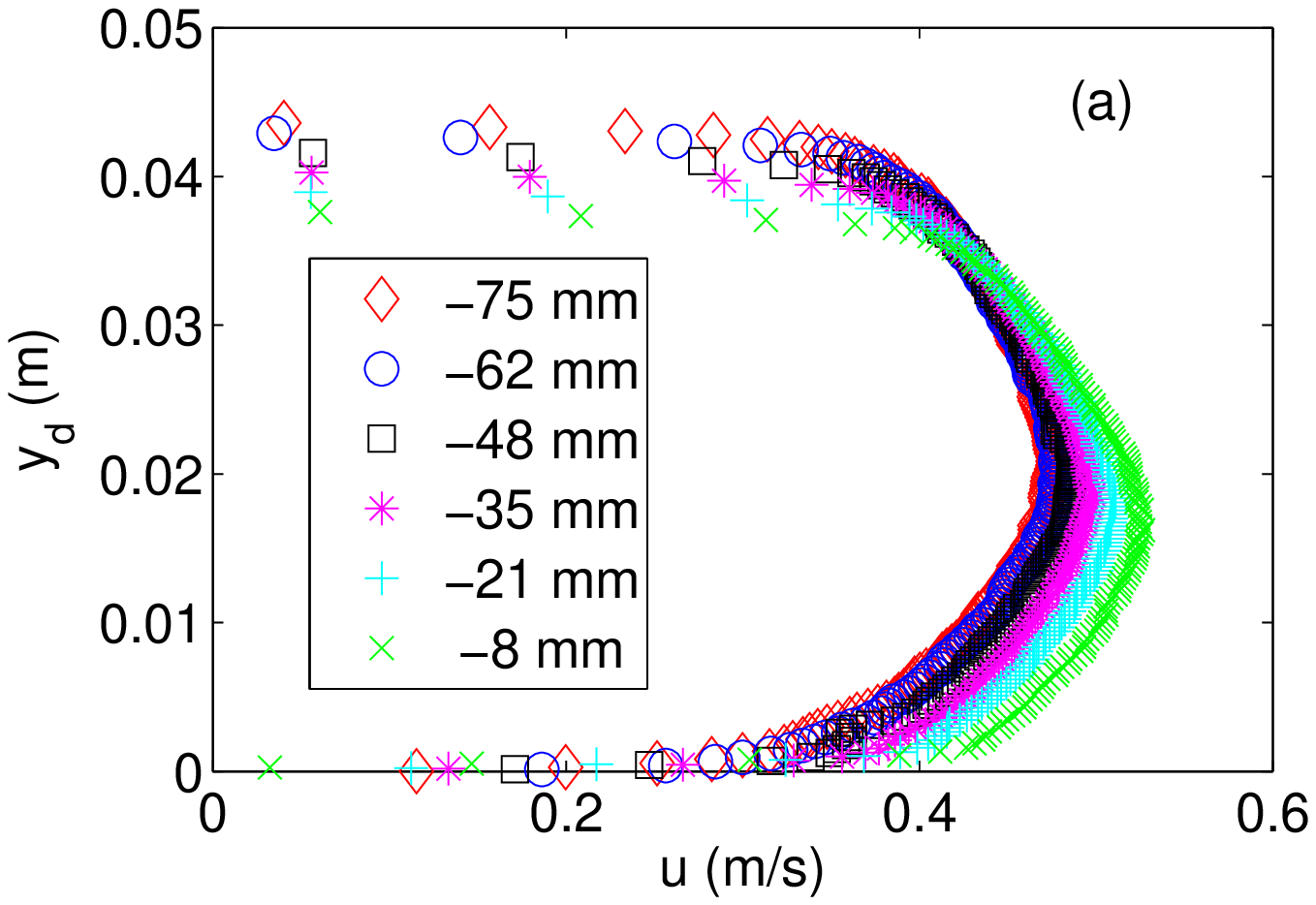}\\
	\includegraphics[scale=0.5]{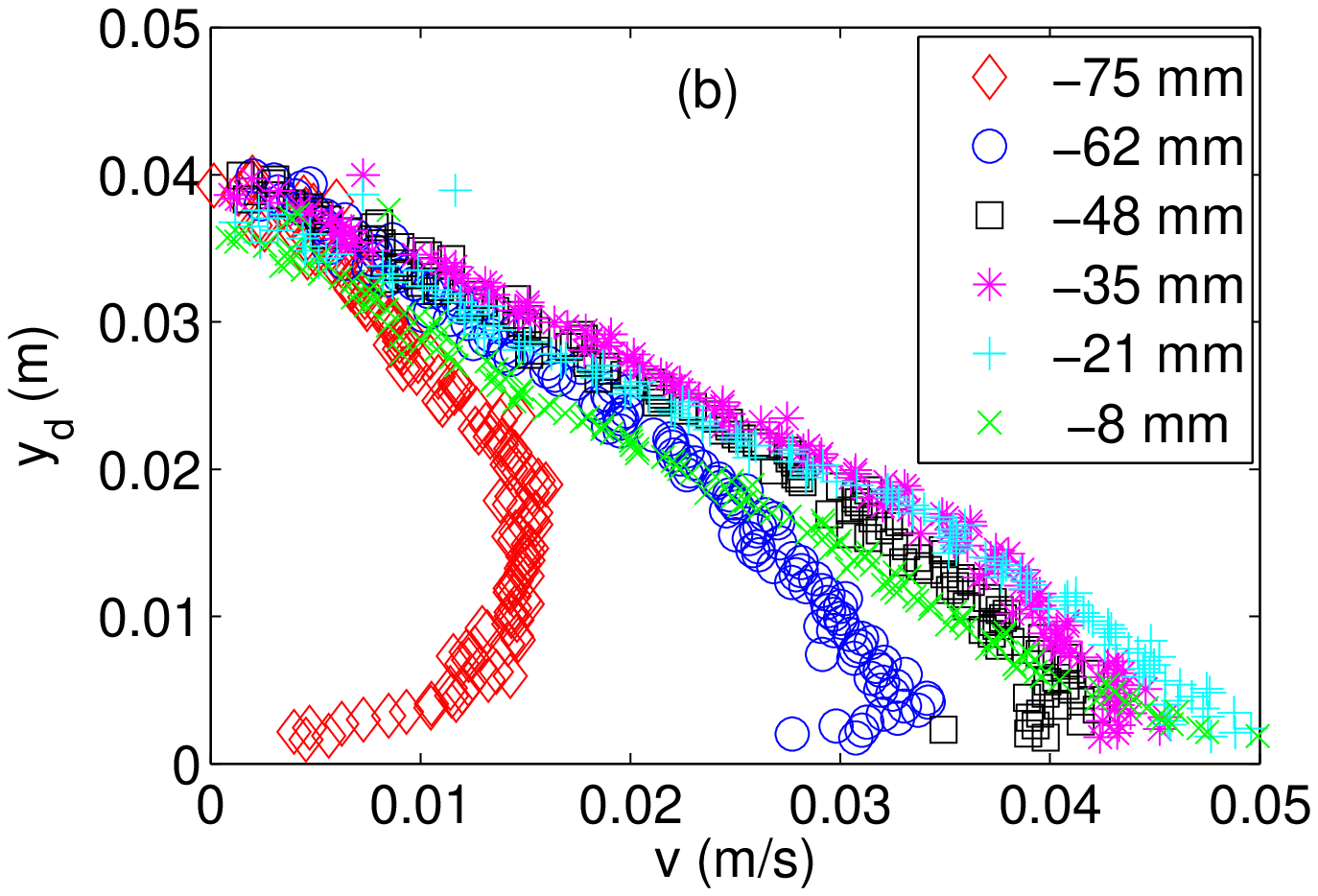}
	\end{tabular} 
\end{center}
\caption{Mean velocity profiles upstream of the ripple crest: (a) displaced vertical coordinate $y_d$ versus the longitudinal mean velocity $u$; (b) displaced vertical coordinate $y_d$ versus the vertical mean velocity $v$. $Re=3.5\cdot 10^4$.}
\label{fig:u_v_mean_ripple}
\end{figure}

\begin{figure}
\begin{center}
	\begin{tabular}{c}
	\includegraphics[scale=0.5]{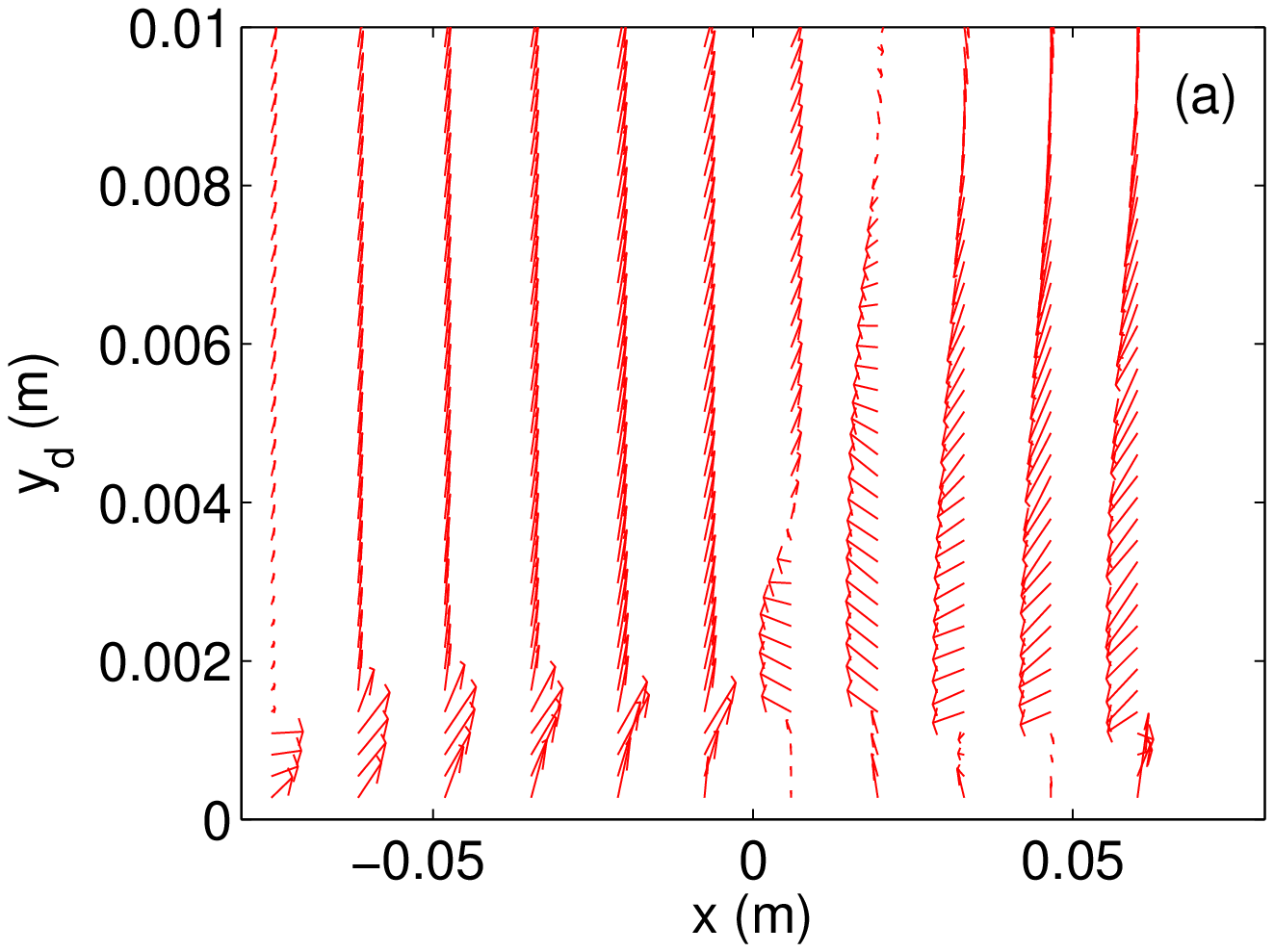}\\
	\includegraphics[scale=0.5]{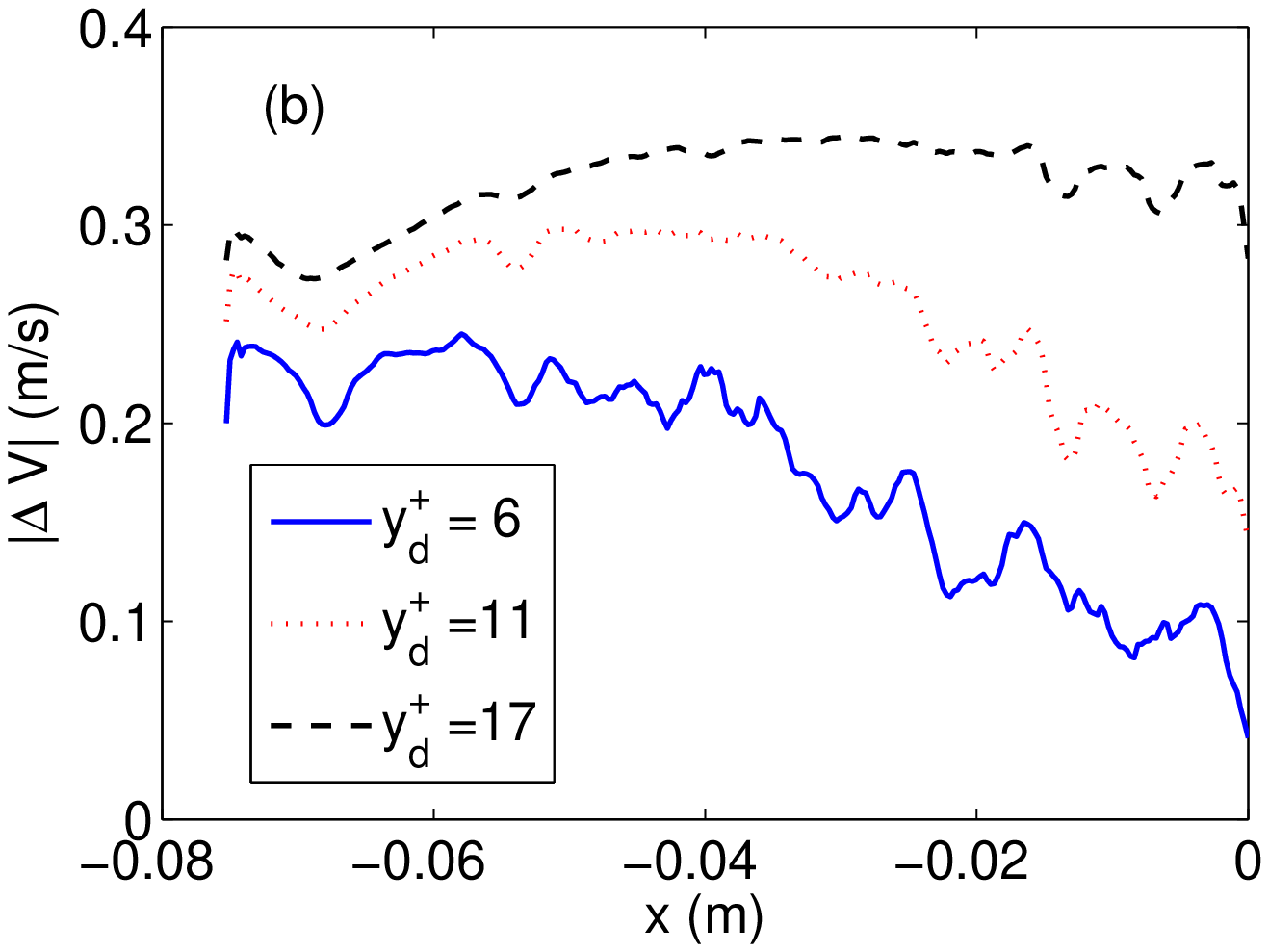}
	\end{tabular} 
\end{center}
\caption{(a) Some perturbation profiles of the mean velocities $\Delta\vec{V}$. (b) Longitudinal evolution of $ |\Delta\vec{V}|$ for three different vertical positions: $y^+_d=6$, $11$ and $17$. $Re=3.5\cdot 10^4$.}
\label{fig:delta_V}
\end{figure}

\begin{sloppypar}
Figure \ref{fig:u_v_mean_ripple} presents the mean velocity profiles upstream of the ripple crest. Figure \ref{fig:u_v_mean_ripple}(a) presents the displaced vertical coordinate $y_d$ versus the longitudinal mean velocity $u$ and shows a convective acceleration of $u$ towards the ripple crest. The degree of acceleration varies with $y_d$ and therefore the longitudinal position where the maximum of $u$ is reached for each value of $y_d$ is not clear in Fig. \ref{fig:u_v_mean_ripple}a. This is presented in more detail in Fig. \ref{fig:delta_V}b.
\end{sloppypar}

Figure \ref{fig:u_v_mean_ripple}(b) presents the displaced vertical coordinate $y_d$ versus the vertical mean velocity $v$. The values of $v$ are one order of magnitude smaller than that of $u$ as expected from the dimensional analysis of the perturbation. The profiles show that $v$ increases from zero at the ripple surface, reaches a maximum and decreases to zero as $y_d$ approaches the upper wall. At $x=-75mm$, the maximum occurs around the center of the channel. Towards the crest, the maximum becomes closer to the ripple surface ($y_d\approx 2mm$). Longitudinally, the maximum of $v$ increases monotonically as the flow approaches the crest.

The perturbation field can be defined as the difference between the flow over the ripple and that over a flat wall, having as reference the solid walls \cite{Jackson_Hunt}. For the mean velocity

\begin{equation}
\Delta\vec{V}(y_d)=\vec{V}(y_d)-\vec{V}_{0}(y_d)
\label{v_perturbed}
\end{equation}

\noindent where $\vec{V}_{0}(y_d)\,=\,u_0(y)\vec{i}$ and $u_0$ is obtained from Eq. \ref{mean_velocity}. The $y_d$ values of each $\vec{V}(y_d)$ profile may differ because the PIV computations employed a Cartesian grid which did not coincide with the ripple inclination. In order to obtain a regular grid for the $\Delta\vec{V}(y_d)$ fields, the $\vec{V}(y_d)$ profiles were interpolated whenever the corresponding values of $y_d$ were not coincident with a pre-determined mesh. The perturbation field can point subtle differences between flows. This is presented in Fig. \ref{fig:delta_V}.

Figure \ref{fig:delta_V}a presents some perturbation profiles of the mean velocities $\Delta\vec{V}(y_d)$ along the ripple. Although the spatial resolution corresponds to $512$ profiles, Fig. \ref{fig:delta_V}a presents only $11$ profiles for a better visualization of the perturbation field. Downstream of the crest, the flow detaches and a recirculation bubble is formed. This is shown in the figure as a region of negative $u$ values, localized within $y_d<8mm$.

Upstream of the crest, the perturbation is localized in the region $y_d<2mm$. If the displaced vertical coordinate is normalized by the viscous length of the unperturbed flow, one obtains $y_d^+=y_du_{*,0}/\nu <43$. This region corresponds to the buffer and viscous sublayers of the unperturbed boundary layer, where the viscous shear stress is very important \cite{Schlichting_1}. In addition, when the water flow occurs in the presence of a granular bed, the transport of grains as bed load takes place very often in the region $y_d^+<20$. Bed load can be defined as a mobile layer of grains rolling and sliding over a fixed bed \cite{Raudkivi_1}.

One of the objectives of this paper is to shed light on the role of the water stream in the formation of sand ripples. As the perturbation and the granular transport (if present) are localized in the region close to the ripple surface, we investigate next the flow in this region. 

Figure \ref{fig:delta_V}b presents the longitudinal evolution of $|\Delta\vec{V}|$ in the $y_d^+<20$ region. Due to the PIV spatial resolution, the presented $|\Delta\vec{V}|$ values correspond to three different vertical positions, $y^+_d=6$, $11$ and $17$. Although the noise is significant, the experimental data show that the perturbation velocity tends to decrease as the flow approaches the crest. Figure \ref{fig:delta_V}b shows that the maximum of the perturbation velocity occurs upstream of the ripple crest, and that the longitudinal distance between the maximum perturbation and the crest decreases as $y_d$ increases.

\begin{figure}
  \begin{center}
    \includegraphics[width=0.80\columnwidth]{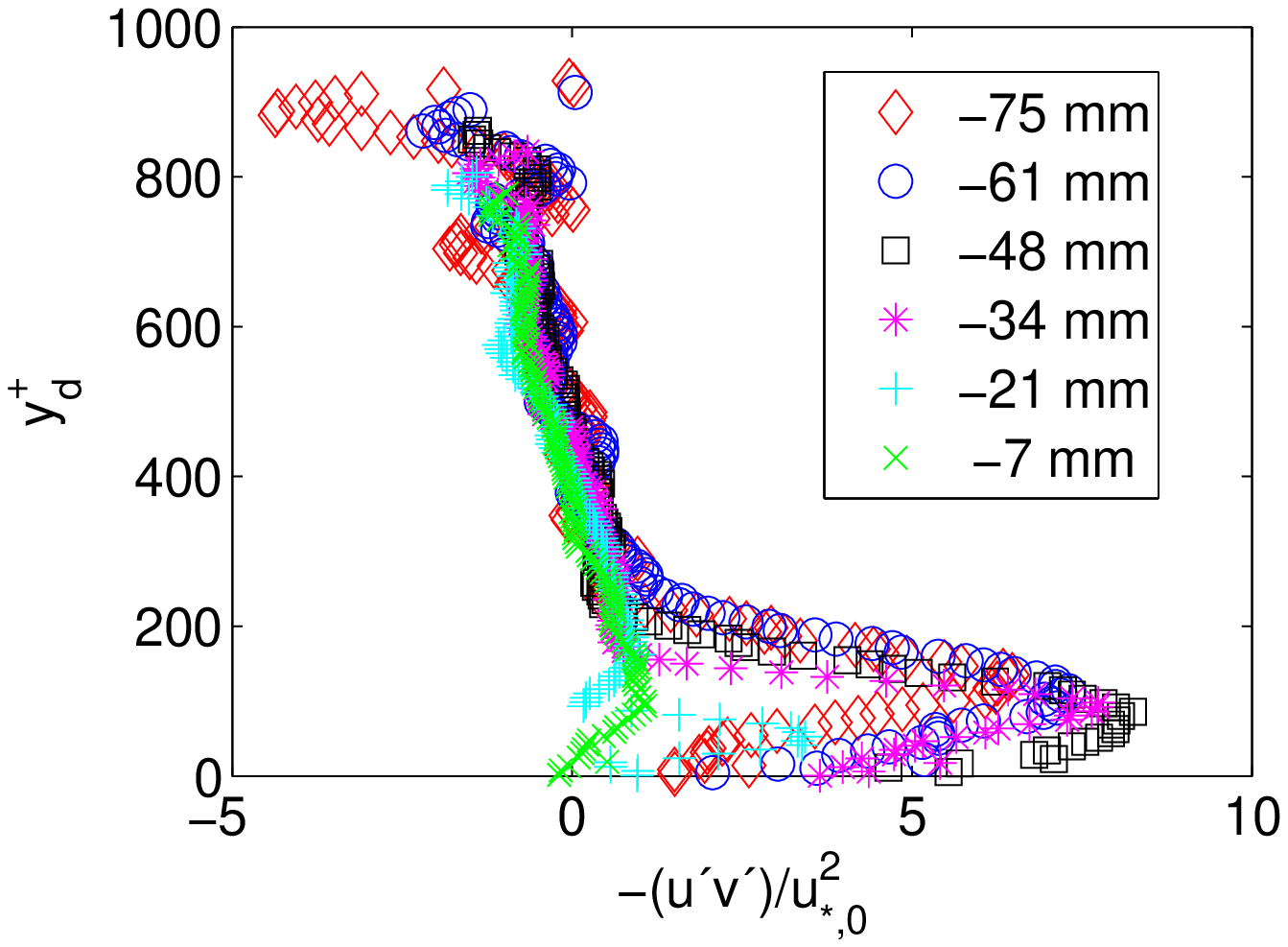}
    \caption{Some profiles of the Reynolds stress $xy$ component in dimensionless form upstream of the ripple crest: $y_d^+$ versus $-\overline{u'v'}/(u_{*,0}^2)$. $Re=3.5\cdot 10^4$.}
    \label{fig:reynolds_ripple}
  \end{center}
\end{figure}

Figure \ref{fig:reynolds_ripple} presents some profiles of the $xy$ component of the Reynolds stresses in dimensionless form, $y_d^+$ versus $-\overline{u'v'}/(u_{*,0}^2)$, upstream of the ripple crest. In order to decrease the noise, the obtained $-\overline{u'v'}$ profiles were averaged by a sliding window process over the closest $9$ points. Figure \ref{fig:reynolds_ripple} shows that the Reynolds stress $-\overline{u'v'}$ is perturbed in the $50<y_d^+<250$ region, that corresponds to the overlap sublayer of the unperturbed boundary layer \cite{Schlichting_1}. If the flow is in local equilibrium in the $y_d^+<250$ region, the shear stress on the surface shall scale with $-\overline{u'v'}$ and therefore the longitudinal evolution of the latter is of importance. Figure \ref{fig:reynolds_ripple} also shows that, longitudinally, the perturbation of $-\overline{u'v'}$ decreases near the crest.  

To understand if the water flow is an unstable mechanism and if the flow is in local equilibrium within $0<y_d^+<250$, the viscous shear stress on the ripple surface $\tau_{visc}$ was evaluated as

\begin{equation}
\tau_{visc}\,=\,\mu\left(\frac{\partial u_{\theta}}{\partial y_{d, \theta}} + \frac{\partial v_{\theta}}{\partial x_{\theta}}\right)\,\approx\,\mu \frac{\partial u_{\theta}}{\partial y_{d, \theta}}
\label{visc_shear}
\end{equation}

\noindent where $u_{\theta}$ and $v_{\theta}$ are, respectively, the aligned and perpendicular components of the mean velocity with respect to the ripple surface, $y_{d, \theta}$ is a displaced coordinate perpendicular to the ripple surface and $x_{\theta}$ is the coordinate aligned with the ripple surface. The derivative was evaluated from a first order Taylor expansion. This component of the viscous stress is the responsible for the transport of grains because any grain transported as bed load shall roll or slide over the ripple surface in a direction aligned with the mean flow.

\begin{figure}
\begin{center}
	\begin{tabular}{c}
	\includegraphics[scale=0.5]{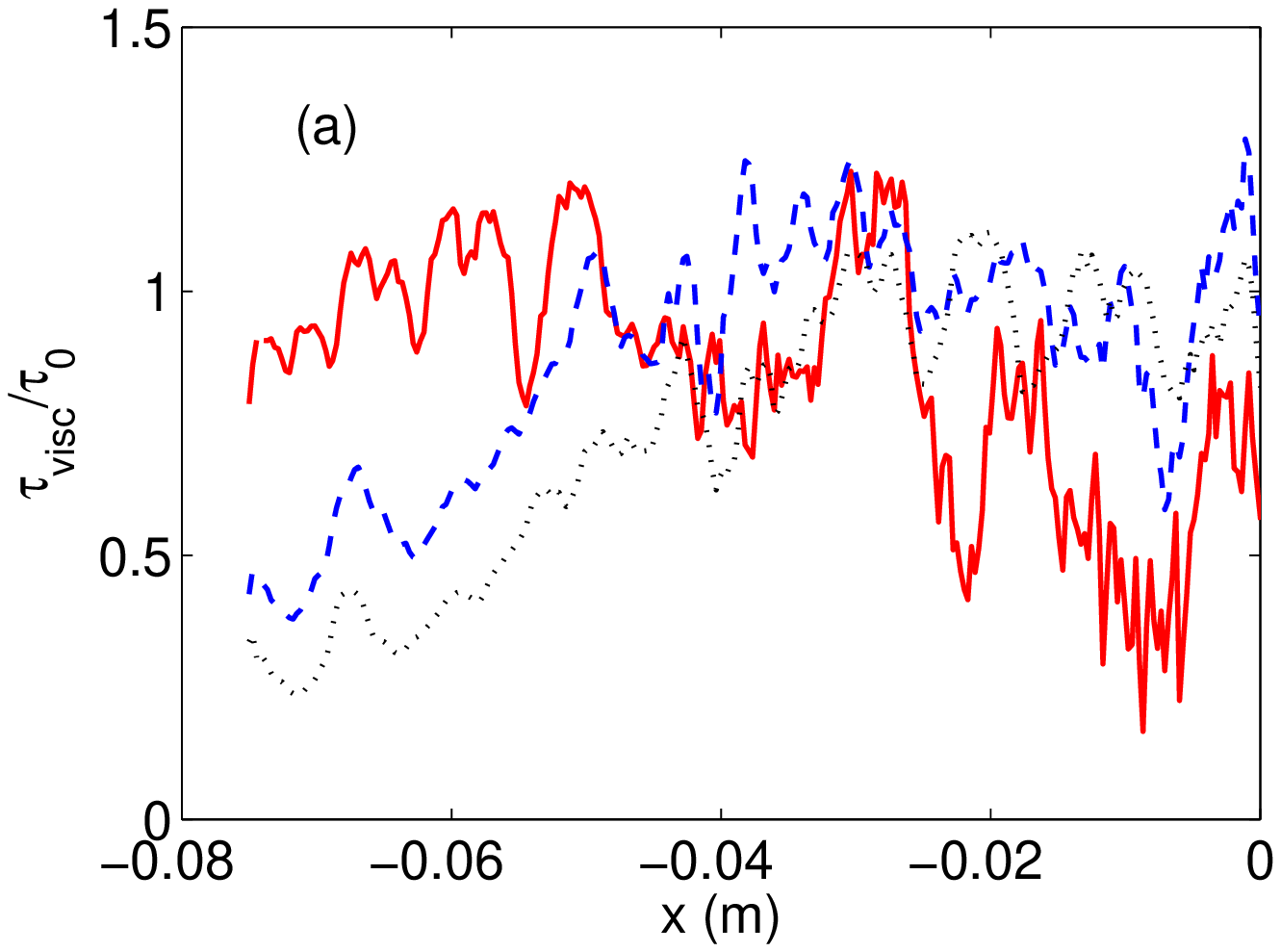}\\
	\includegraphics[scale=0.5]{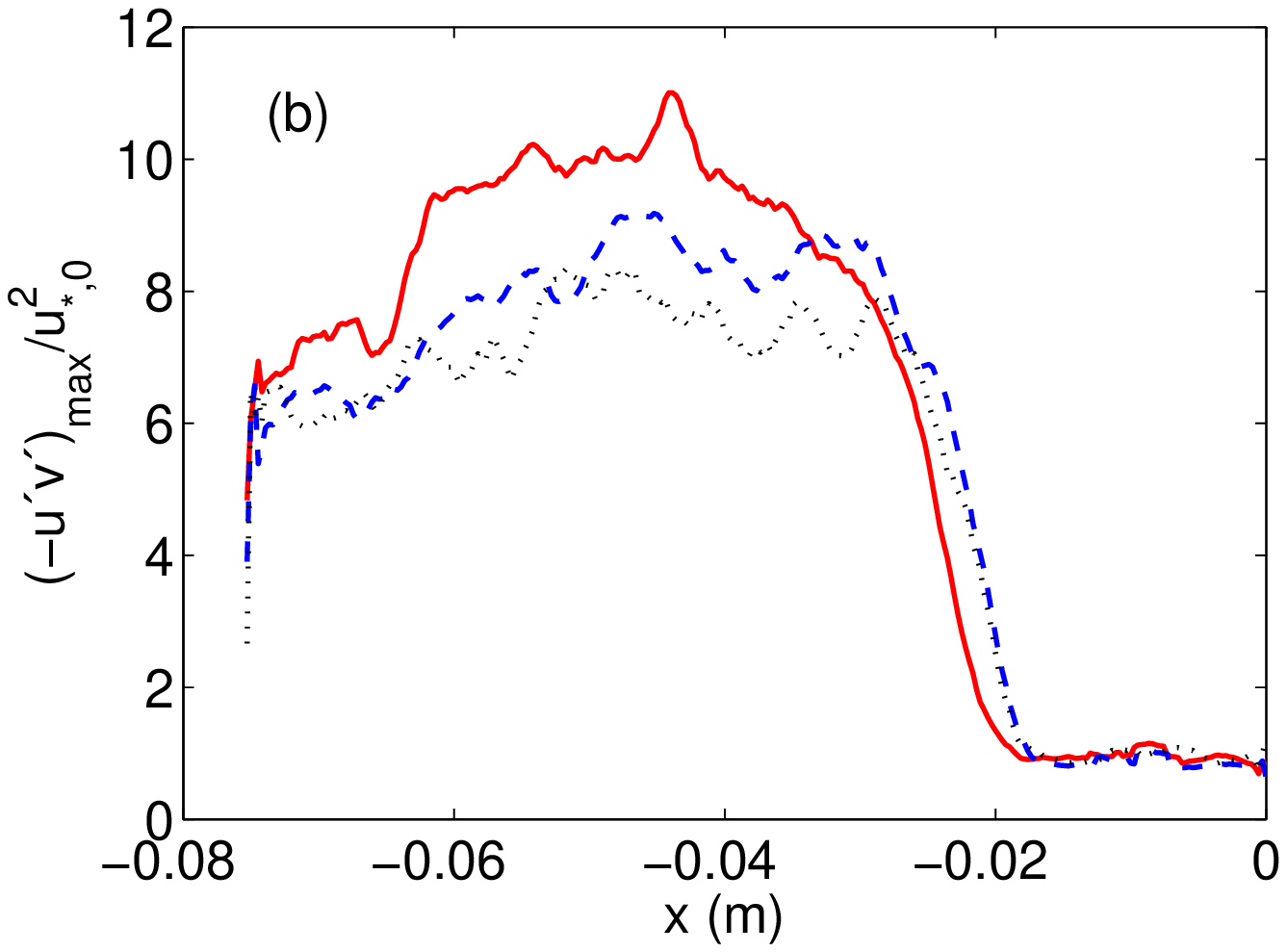}
	\end{tabular} 
\end{center}
\caption{(a) Normalized surface viscous stress $\tau_{visc}/\tau_0$ as a function of the longitudinal position $x$. (b) Maximum normalized Reynolds stress $\left( -\overline{u'v'}\right)_{max}/(u_{*,0}^2)$ as a function of the longitudinal position $x$. The continuous, dashed and dotted lines correspond to $Re=1.8\cdot 10^4$, $2.6\cdot 10^4$ and $3.5\cdot 10^4$, respectively.}
\label{fig:shear_stress}
\end{figure}

Figure \ref{fig:shear_stress}a presents the evolution along the ripple of the surface viscous stress $\tau_{visc}$ normalized by the unperturbed shear stress $\tau_0=\rho u_{*,0}^2$. In this figure, the continuous, dashed and dotted lines correspond to $Re=1.8\cdot 10^4$, $2.6\cdot 10^4$ and $3.5\cdot 10^4$ ($\overline{U}=0.2m/s$, $0.3m/s$ and $0.4m/s$), respectively. From this figure, it seems that the maximum of the surface stress occurs upstream of the ripple crest. However, this cannot be asserted because of the relatively high noise in the data.

If the flow is in local equilibrium within $0<y_d^+<250$, the surface shear stress shall be proportional to the maximum of $-\overline{u'v'}$, which in its turn scales with $u_*^2$. Figure \ref{fig:shear_stress}b presents the longitudinal evolution of the the maximum of the Reynolds stress $\left( -\overline{u'v'}\right)_{max}$ normalized by $u_{*,0}^2$. The lines are the same as in Fig. \ref{fig:shear_stress}a.

\begin{sloppypar}	
Figure \ref{fig:shear_stress}b shows that for all cases $\left( -\overline{u'v'}\right)_{max}$ reaches high values of approximately $8u_{*,0}^2$ in the $x\lesssim -0.025m$ region (upstream of the ripple crest), with a peak at $x\approx -0.05m$ (for the triangular shape, at $x/L\approx 1$). Within $-0.025m\lesssim x\lesssim -0.020m$, $\left( -\overline{u'v'}\right)_{max}$ decreases rapidly to $\left( -\overline{u'v'}\right)_{max}\approx u_{*,0}^2$, and it remains stable at this value in the $-0.020m\lesssim x\leq 0m$ region. The upstream position of $\left( -\overline{u'v'}\right)_{max}$ agrees with a possible upstream shift of $\tau_{visc}$. However, in terms of magnitude, there is a large difference between them, with $\tau_{visc}/\left( -\rho\overline{u'v'}\right)_{max} = O(0.1)$ in the $x\lesssim -0.025m$ region. This indicates that, for the water flow over the triangular ripple, the region corresponding to the overlap sublayer of the unperturbed flow is not in local equilibrium with the lower sublayers.
\end{sloppypar}	

The absence of local equilibrium in the perturbed $-\overline{u'v'}$ region means that the mean velocity profiles are no longer perfectly logarithmic. In the case of aeolian dunes, the existence of a logarithmic sublayer in the perturbed flow has been discussed \cite{Bagnold_1}, some authors considering that this layer exists \cite{Parteli_2}. In fact, if not logarithmic, deviations from the logarithmic profiles are subtle in the aeolian case given the large longitudinal scale of the problem: the wavelength of aeolian dunes is $1000$ larger than that of aquatic ripples, so that the ratio between the vertical scales (thickness of the inner sublayers) and the longitudinal wavelength is, at least, $100$ smaller in the aeolian case. In these conditions, a local-equilibrium inner region can exist.

\begin{sloppypar}
In the case of aquatic ripples, local-equilibrium conditions may not occur. However, instability analyses usually employ perturbation expressions obtained from the local-equilibrium assumption. For some aquatic cases, other expressions should be proposed.
\end{sloppypar}

If the shear stress on the ripple surface is assumed to vary with $-\overline{u'v'}$ (even if the flow is not in local equilibrium) then the maximum stress occurs upstream of the ripple crest, what seems to agree with $\tau_{visc}$. In this case, the perturbed fluid flow is an unstable mechanism in what concerns the formation of sand ripples. Although the precise magnitude of the surface shear stress cannot be determined due to high uncertainties in $\tau_{visc}$ as well as to the absence of local-equilibrium conditions, the present findings contribute to the understanding of the unstable role of the water flow.

\section{CONCLUSIONS}
\label{section:conclusions}

This paper investigated the perturbation of a liquid turbulent boundary layer by a two-dimensional ripple in the hydraulic smooth regime. Measurements of a closed-conduit water flow perturbed by a model ripple of triangular shape were made by PIV and the obtained mean and turbulent fields were compared with the unperturbed flow fields. Some known characteristics of the boundary-layer perturbation by a low hill were confirmed by the experimental results: a recirculation bubble is formed downstream of the ripple crest and the perturbation is localized in a region close to the ripple surface ($y_d^+\lesssim 40$). However, some characteristics not considered until now, or for which a consensus has not been achieved, had their importance shown by the presented results.

Due to the relatively large ratio between the vertical and longitudinal flow scales (when compared to the aeolian case), the inner regions of the perturbed flow are not in local-equilibrium for some aquatic ripples. This means that asymptotic expressions for the boundary-layer perturbation based on local-equilibrium conditions must be used with care in the case of aquatic sand ripples of triangular shape, specially if Reynolds numbers are $Re<10^5$. To the authors' knowledge, this is the first time that absence of local-equilibrium conditions is proposed for the flow over a triangular ripple in moderate Reynolds numbers ($10000 < Re < 50000$).

\begin{sloppypar}	
The maximum of the shear stress on the ripple surface occurs upstream of the ripple crest. For the triangular ripple, the maximum surface shear stress was found to occur at $50\%$ of the ripple length (in this case, $\approx L$). However, more experiments employing other shapes as well as increased PIV spatial resolutions are necessary to quantify the characteristic length of the shift. This shift is necessary to explain the instabilities giving rise to aquatic ripples in the hydraulic smooth regime. The present experimental findings are of importance to understand the formation, the evolution and the stability of bedforms under turbulent liquid flows.
\end{sloppypar}

\begin{acknowledgements}
The authors are grateful to Petrobras S.A. (contract number 0050.0045763.08.4). Erick M. Franklin is grateful to FAEPEX/UNICAMP (conv. 519.292, project 1435/12) and to FAPESP (contract number 2012/19562-6).
\end{acknowledgements}


\bibliography{references}
\bibliographystyle{spphys}

\end{document}